\documentclass[aps,prb,twocolumn]{revtex4-1}

\usepackage{amsmath}
\usepackage{graphicx}
\usepackage{dcolumn}
\usepackage{bm}
\usepackage{mathrsfs}
\usepackage{mhchem}
\usepackage{subfigure}
\usepackage{color}
\draft

\begin{document}

\title{A microscopic model for chemically-powered Janus motors}

 \author{Mu-Jie Huang}
 \email{mjhuang@chem.utoronto.ca}
 \author{Jeremy Schofield}
 \email{jmschofi@chem.utoronto.ca}
 \author{Raymond Kapral}
 \email{rkapral@chem.utoronto.ca}
 \affiliation{Chemical Physics Theory Group, Department of Chemistry, University of Toronto, Toronto, Ontario M5S 3H6, Canada}

\begin{abstract}
Very small synthetic motors that make use of chemical reactions to propel themselves in solution hold promise for new applications in the development of new materials, science and medicine. The prospect of such potential applications, along with the fact that systems with many motors or active elements display interesting cooperative phenomena of fundamental interest, has made the study of synthetic motors an active research area. Janus motors, comprising catalytic and noncatalytic hemispheres, figure prominently in experimental and theoretical studies of these systems. While continuum models of Janus particle systems are often used to describe motor dynamics, microscopic models that are able to account for intermolecular interactions, many-body concentration gradients, fluid flows and thermal fluctuations provide a way to explore the dynamical behavior of these complex out-of-equilibrium systems that does not rely on approximations that are often made in continuum theories. The analysis of microscopic models from first principles provides a foundation from which the range of validity and limitations of approximate theories of the dynamics may be assessed.  In this paper, a microscopic model for the diffusiophoretic propulsion of Janus motors, where motor interactions with the environment occur only through hard collisions, is constructed, analyzed and compared to theoretical predictions. Microscopic simulations of both single-motor and many-motor systems are carried out to illustrate the results.
\end{abstract}
\maketitle

\section{Introduction}

Synthetic chemically propelled motors of various shapes and sizes have been the object of a considerable amount of research.\cite{wangbook:13,SenRev:13,kapral:13,sanchez:14} Interest in such self-propelled objects derives both from their potential uses, such as nanoscale cargo delivery vehicles, and because, like all active matter operating under nonequilibrium conditions, they display phenomena that differ from those in equilibrium systems. Although such motors may utilize chemical activity in different ways to produce directed motion, this paper concerns motors that operate by diffusiophoretic mechanisms. In self-diffusiophoresis, asymmetric catalytic activity on the motor produces a concentration gradient in chemical species which gives rise to a force that is responsible for directed motion. Of the possible motor geometries, perhaps the simplest is a spherical Janus particle where one hemisphere catalyzes the conversion of fuel to product while the other hemisphere is chemically inactive. Janus particles may be readily made in the laboratory, the macroscopic theory for their self-diffusiophoretic propulsion is well developed and the spherical symmetry of a single Janus motor simplifies the theoretical calculations. For these reasons they have been the subjects of extensive experimental and theoretical study.~\cite{golestanian-1:07,showalter:10,palacci:10,Ebbens2012,Baraban2012,debuyl:13}

The collective dynamics seen in systems containing many motors depends on the interactions among motors as well as the propulsion properties of the individual motor constituents.~\cite{Ramaswamy_10,Vicsek_Zafeiris_12,bocquet:12,Wang_etal_13,Cates_Tailleur_13,Marchetti_etal_13,Pohl_Stark_14,Saha_etal_14,Bialke_Speck_Lowen_15,Magistris_Marenduzzo_15,Yadav_etal_15,Elgeti_Winkler_Gompper_15} These interactions among motors can arise from distinct origins, including direct intermolecular interactions as well as coupling through hydrodynamic flow fields and chemical gradients. Furthermore, small motors are subject to strong thermal fluctuations, which must also be included in any theoretical description of the dynamics. Continuum models supplemented with Langevin forces may be used to study the behavior of these systems; however, there is a need for microscopic theories for the dynamics of active systems for a number of reasons. Studies of chemically-powered motors are being extended to motors with very small spatial dimensions down to the scale of tens of nanometers or even Angstroms.~\cite{colberg:14,lee:14} On such small length scales the validity and applicability of continuum theories requires reexamination. In addition, by treating direct motor interactions from first principles, microscopic theories will automatically account for many-body hydrodynamic interactions and chemical gradients on both large and small scales. These desirable features are achieved at the cost of having to explicitly treat the dynamics of all constituents of the system, namely the motors, the reactive chemical species and the solvent, at a particle-based level.

Microscopic models have been constructed previously and used to investigate chemically-powered motors.~\cite{Rueckner2007,yuguo:08,Thakur_Kapral_11,debuyl:13,Yang_Wysocki_Ripoll_14,li15,Thakur_15,Fedosov_Sengupta_Gompper_15,JXChen_16} In contrast to these models, the microscopic model described in this paper involves only hard interactions between the Janus motor and solvent species. The model captures all of the essential features of the motor mechanism, is simple to treat theoretically and has the advantage that its dynamics may be simulated efficiently.

The paper is structured as follows: The model for an active Janus motor propelled through a diffusiophoretic mechanism is described in Sec.~\ref{sec:simple-model}. The corresponding continuum theory, which includes a description of how the system is maintained out of equilibrium, is presented in Sec.~\ref{sec:continuum-model}. The simulation method and system parameters are given in Sec.~\ref{sec:micro-sim}. Simulation results for the dynamical properties of a single Janus particle, along with a discussion of the quantities needed to make a comparison with the continuum theory, can be found in Sec.~\ref{sec:one-janus}. Section~\ref{sec:collective} gives a brief description of the behavior of many Janus motors to show that the model is able to describe the important many-body aspects of the collective dynamics. A discussion of the results in the paper are given in Sec.~\ref{sec:conc}.

\section{Microscopic model for Janus motors}\label{sec:simple-model}
In this paper we consider a particle-based microscopic model that combines molecular dynamics for the motor interacting with the solvent, including reactive chemical species, with a coarse-grain description of the interactions among all solvent species. In this model, solvent particles interact periodically only through an effective collision operator described in detail in Sec.~\ref{sec:micro-sim} and otherwise stream freely between collisions with the Janus motor.

\begin{figure}[htbp]
\centering
\resizebox{1.0\columnwidth}{!}{%
\includegraphics{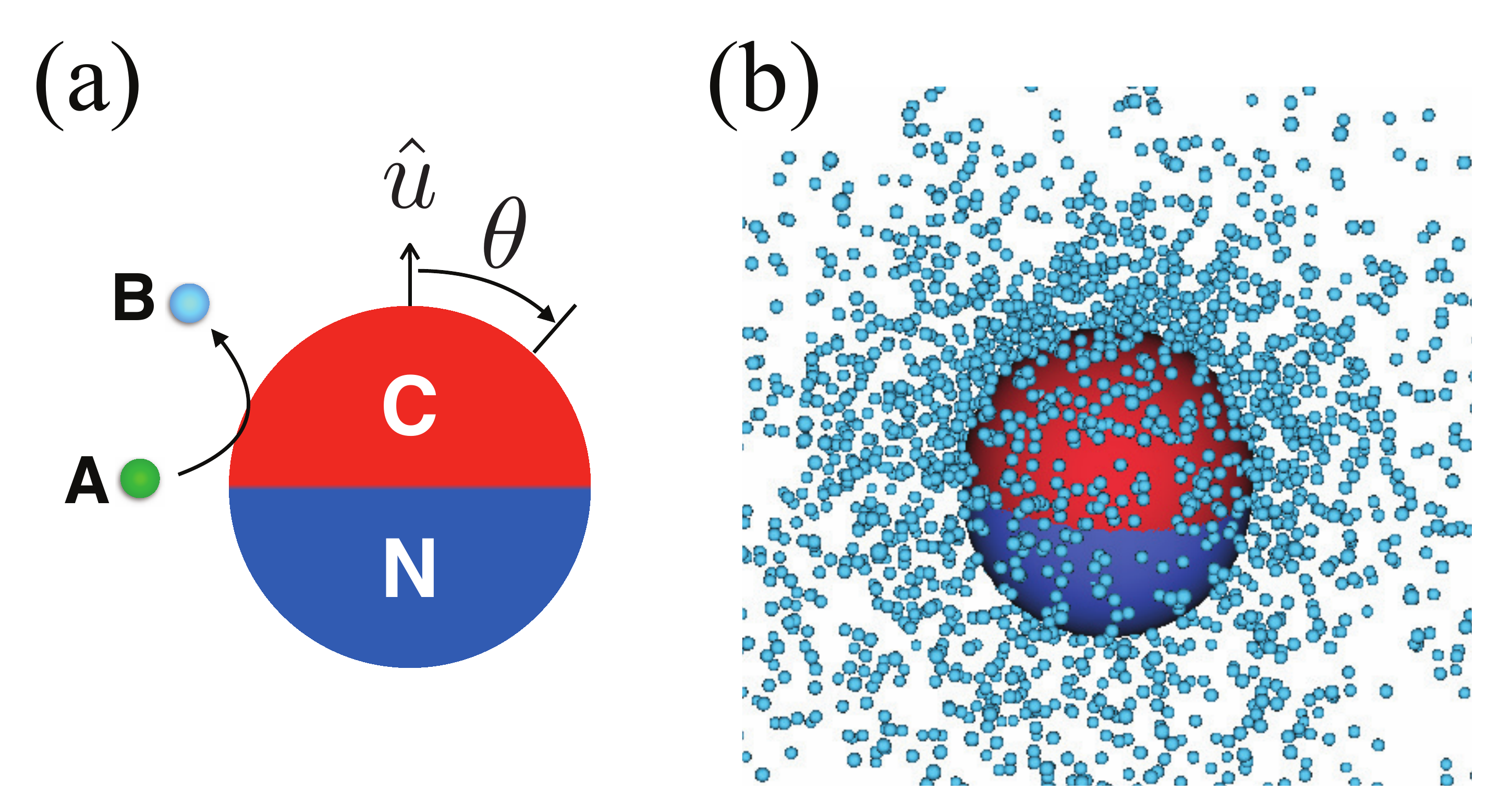} }
\caption{\label{fig:JP}(a) Sketch of the Janus particle comprising catalytic ($C$) and noncatalytic (N) hemispherical surfaces. The chemical reaction, $A \rightarrow B$, occurs on the $C$ surface and converts fuel $A$ particles (green) to product $B$ particles (light blue). The orientation of the Janus particle $\hat{\pmb{u}}$ and the polar angle $\theta$ are indicated. (b) An instantaneous configuration drawn from the simulation of the dynamics of the system shows the distribution of $B$ particles in the vicinity of the Janus particle.}
\end{figure}

Consider a Janus motor of radius $R$ with catalytic and non-catalytic hemispheres, denoted as $C$ and $N$, respectively. As shown in Fig.~\ref{fig:JP}(a), a chemical reaction, $A \rightarrow B$,  takes place on the $C$ hemispherical surface that converts fuel particles $A$ to product particles $B$ and, in the process, produces a concentration gradient of $A$ and $B$ particles in the vicinity of the motor (Fig.~\ref{fig:JP}(b)). We assume that such a reaction occurs whenever an $A$ particle collides with the catalytic hemisphere.

The $A$ and $B$ solvent particles interact with the Janus motor through a variant of hard-sphere collisions in which the solvent particles are allowed to penetrate the Janus sphere but experience modified bounce-back collisions when their distance to the sphere is less than a specified collision radius. More specifically, at a position $r$ from the center of mass of Janus motor, the $A$ and $B$ particles interact with the motor through hard potentials $W_{\alpha J}(r)$,
\begin{equation}
	W_{\alpha J}(r) = \left\{
	\begin{array}{ll}
	\infty\mbox{ },& r < R_{\alpha}\\
%	\\
	0\mbox{ },& r \geq R_{\alpha},
	\end{array}
	\right.
	 \label{eq:hard_potential}
\end{equation}
where $\alpha = A,B$, and $R_{\alpha} \leq R$ is the collision radius for a particle of type $\alpha$ interacting with the motor surface. We let $R$ denote the larger of $R_A$ and $R_B$. The collision radii are chosen so that the quantity $R -{R}_{\alpha}$ is small compared to the motor radius.

The rules that govern the bounce-back collisions with the Janus motor are as follows: Let $\mathbf{r}$ and $\mathbf{v}$ be the position and velocity of the solvent particle of species $\alpha$ with mass $m$, and $\mathbf{r}_J$ , $\mathbf{v}_J$ and $\pmb{\omega}_J$ be the position, linear velocity and angular velocity of the Janus motor with mass $M$ and moment of inertia $I$. The relative position and velocity are defined by $\mathbf{r}^* = \mathbf{r} - \mathbf{r}_J$ and $\mathbf{v}^* = \mathbf{v} - \mathbf{v}_J$.

\begin{figure}[htbp]
\centering
\resizebox{0.7\columnwidth}{!}{%
\includegraphics{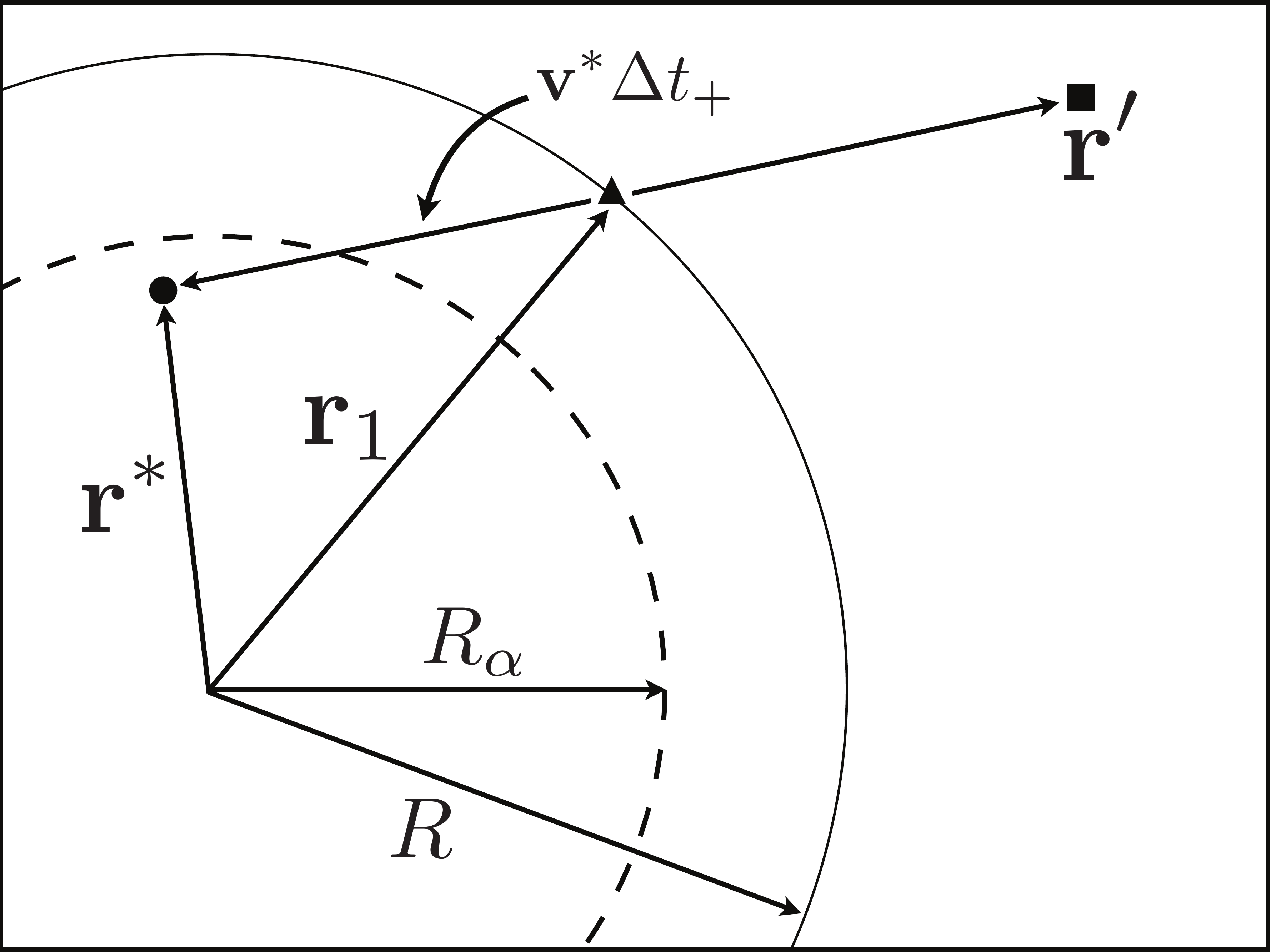}}
\caption{\label{fig:BouncebackDiagram}Application of the bounce-back collision rule for a solvent particle of type $\alpha$. When a solvent particle moves with velocity $\mathbf{v}^*$ toward the Janus particle and finds itself at position $\mathbf{r}^*$ (black dot) inside the reflecting radius $R_{\alpha}$ (dashed circle), a bounce-back collision takes place at the contact position $\mathbf{r}_1$ (black triangle) on the surface of the Janus particle with radius $R$ (solid circle). The travel time from $\mathbf{r}_1$ to $\mathbf{r}^*$ is $\Delta t_+$ and the postcollision position of the solvent particle is $\mathbf{r}'$ (black square).}
\end{figure}

The bounce-back collision dynamics differs for the $A$ and $B$ particles and, referring to Fig.~\ref{fig:BouncebackDiagram}, can be described as follows. During free streaming the instantaneous relative positions of solvent particles of each species $\alpha$ are monitored at each time step $\delta t$ and a bounce-back collision will occur if $(\mathbf{r}^*\cdot\mathbf{v}^*)<0$, and $|r^*| < R_{\alpha}$ so that the particle encountered the collision surface with radius $R_{\alpha}$.
We assume that the solvent and Janus motor exchange momentum at position $\mathbf{r}_1 = R\:\hat{r}_1$ on the surface of the Janus motor, and that $|\mathbf{v}| \gg |\mathbf{v}_J|$ and $|\lambda - (R-R_{\alpha)}|$ is small, where $\lambda$ is the mean free path of the solvent particle, so that the relative velocities are approximately the same at the positions $\mathbf{r}^*$ and $\mathbf{r}_1$. To compute $\mathbf{r}_1$, one needs to find the time $\Delta t$ that it takes the solvent particle to travel from $\mathbf{r}_1$ to $\mathbf{r}^*$. Note that $\mathbf{r}^* = \mathbf{r}_1 + \mathbf{v}^*\Delta t$, which yields $|\mathbf{r}^* - \mathbf{v}^* \Delta t| = |\mathbf{r}_1|$. This travel time is
\begin{equation}
\Delta t_{\pm} = \frac{\mathbf{r}^*\cdot\mathbf{v}^*}{v^{*2}} \pm \frac{1}{v^{*2}} \sqrt{(\mathbf{r}^*\cdot\mathbf{v}^*)^2 - v^{*2}(r^{*2} - R^2)},
\end{equation}
where the solution $\Delta t_+$ is taken, since $\Delta t_-$ is the time it takes the solvent particle to travel to the farther surface of the Janus particle. Therefore, we have
\begin{equation}
\mathbf{r}_1 = \mathbf{r}^* - \mathbf{v}^* \Delta t_+.
\end{equation}

At position $\mathbf{r}_1$, the velocity of the solvent particle, which is treated as a spin-less point particle, relative to the moving and rotating Janus motor surface is
\begin{equation}
\tilde{\mathbf{v}} = \mathbf{v}^* - \pmb{\omega}_J \times \mathbf{r}_1 = \tilde{\mathbf{v}}_n + \tilde{\mathbf{v}}_t ,
\end{equation}
where $\tilde{\mathbf{v}}_n = \hat{\mathbf{r}}_1(\hat{\mathbf{r}}_1\cdot \tilde{\mathbf{v}})$ and $\tilde{\mathbf{v}}_t = \tilde{\mathbf{v}} - \tilde{\mathbf{v}}_n$ are the normal and tangential components of $\tilde{\mathbf{v}}$, respectively. The momentum exchange during each collision is given by
\begin{equation}
\Delta \mathbf{p} = \Delta\mathbf{p}_n + \Delta\mathbf{p}_t = -\mu \Delta \tilde{\mathbf{v}}_n - \frac{\mu I}{I + \mu R^2} \Delta \tilde{\mathbf{v}}_t,
\end{equation}
where $\mu = mM/(m+M)$ is the reduced mass.\cite{Whitmer_Luijten_10,Yang_Wysocki_Ripoll_14} After a collision, the relative velocity is completely reversed, $\tilde{\mathbf{v}}' = - \tilde{\mathbf{v}}$, and the velocity changes in the normal and tangential directions are $\Delta \tilde{\mathbf{v}}_n = \tilde{\mathbf{v}}_n' - \tilde{\mathbf{v}}_n = -2\:\tilde{\mathbf{v}}_n$ and $\Delta \tilde{\mathbf{v}}_t = \tilde{\mathbf{v}}_t' - \tilde{\mathbf{v}}_t = -2\:\tilde{\mathbf{v}}_t$, respectively. Then, the post-collision linear and angular velocities are given by
\begin{eqnarray}
\mathbf{v}' &=& \mathbf{v} - \Delta\mathbf{p}/m, \\
\mathbf{v}_J' &=& \mathbf{v}_J + \Delta\mathbf{p}/M, \quad  \pmb{\omega}_J' =
\pmb{\omega}_J - (\mathbf{r}_1\times\Delta\mathbf{p})/I.\nonumber
\end{eqnarray}
These collision rules conserve the energy as well as the total linear and angular momentum of the system.
After the collision at the surface of the Janus motor, the post-collision position of the solvent particle is taken to be $\mathbf{r}'$,
\begin{equation}
\mathbf{r}' = \mathbf{r} - 2\:\mathbf{v}^* \Delta t_+.
\label{eq:r'}
\end{equation}

\begin{figure}[htbp]
\centering
\resizebox{0.6\columnwidth}{!}{%
\includegraphics{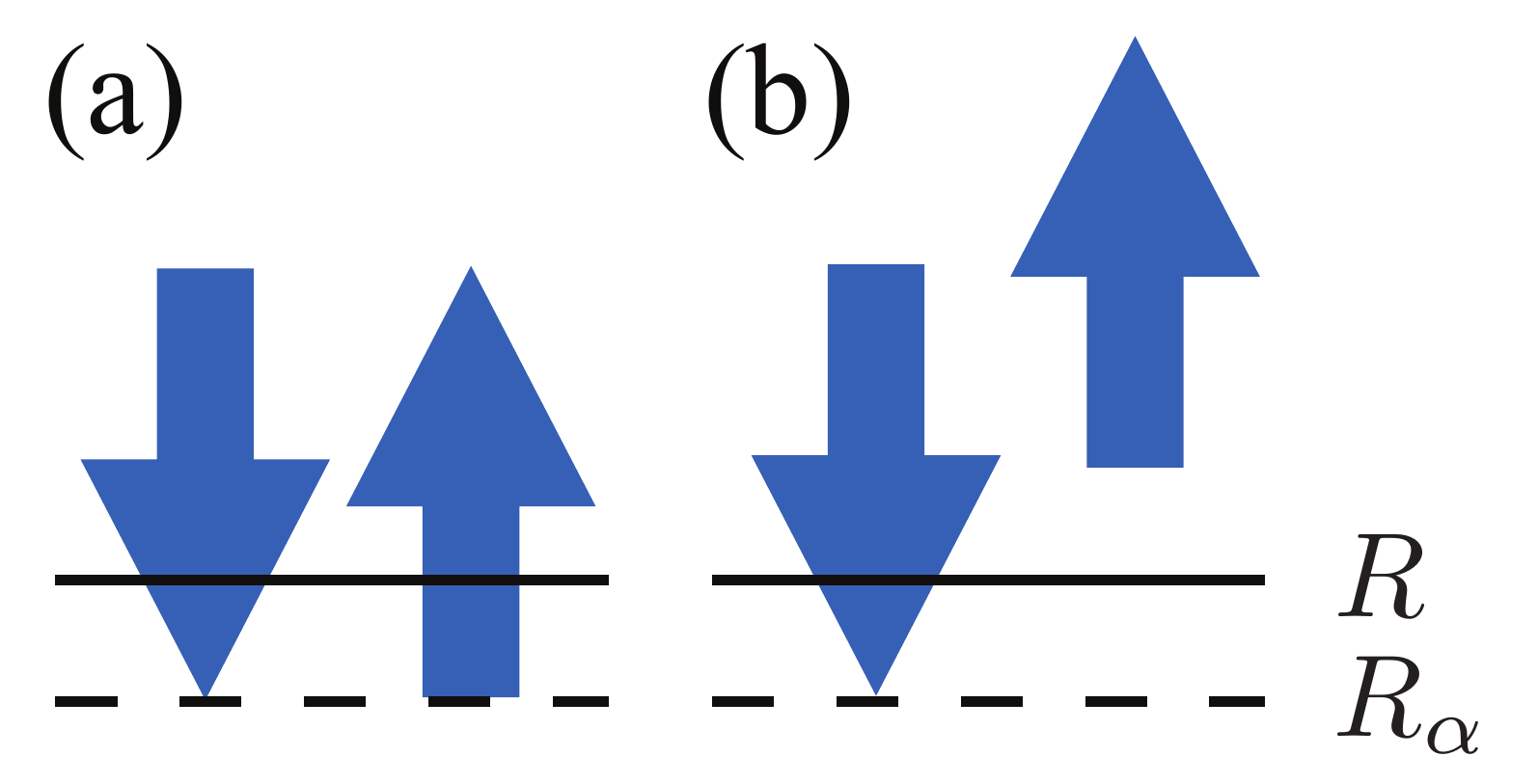}}
\caption{\label{fig:effective_radius}The collision radius for (a) conventional bounce-back rule at $R_{\alpha}$ and (b) modified bounce-back rule at $R$. The incoming and outgoing solvent particles are indicated as blue arrows. }
\end{figure}

Figure~\ref{fig:effective_radius} (a) shows the conventional bounce-back rule, where $\mathbf{r}_1 = R_{\alpha} \hat{r}_1$, and the particle velocity is reversed upon collision with the surface at $R_\alpha$. In the presence of soft repulsive interactions particles are repelled from the surface and to mimic this effect in the modified bounce-back model the position of a solvent particle is shifted  according to Eq.~(\ref{eq:r'}), and so is the outgoing particle flux (Fig.~\ref{fig:effective_radius} (b)). While both simple and modified bounce-back rules give rise to directed motion, in the modified bounce-back collision rule the solvent particles are forced to leave the Janus motor after collision thereby incorporating an effective repulsion in the dynamics. In general, depending on the nature of intermolecular interactions between the surface of the Janus motor and solvent species, one may chose various bounce-back rules that give rise to directed motion. In these bounce-back collision models, since the $A$ and $B$ species have different collision cross sections, $\pi R_\alpha^2$ with the Janus particle, and a concentration gradient of these species is present, a body force on the motor is produced which leads to directed motion by the diffusiophoretic mechanism.

\section{Continuum model for Janus particle velocity}\label{sec:continuum-model}

Theoretical predictions of the diffusiophoretical motion of Janus motors based on continuum theory in a low P{\'{e}}clet number regime have been developed previously.~\cite{anderson:89,golestanian:05} The continuum treatment assumes that the fluid and reactive species concentration fields are described by the Stokes and reaction-diffusion equations, respectively.  The fluid velocity field satisfies stick boundary conditions on the surface of the motor, and the concentration fields satisfy reactive ``radiation" boundary conditions on the catalytic part of the Janus motor. Since the reactive chemical species have different interactions with the Janus motor, the self-generated inhomogeneous concentration field gives rise to a net body force on the motor, which, in turn, produces a fluid flow in the boundary layer around the Janus motor within which forces act. The resulting fluid velocity field at the outer edge of the boundary layer is the slip velocity, and this slip velocity provides a boundary condition for the solution of the Stokes equation and thereby determines the velocity field outside the boundary layer that accompanies the Janus motor motion. In this continuum theory, the propulsion velocity of the Janus motor along its symmetry axis, $\hat{\pmb{u}}$, can be calculated from the surface average of the slip velocity, ${V}_u = -\langle\hat{\pmb{u}}\cdot \mathbf{v}^{(s)}\rangle_S$, where $\langle \cdots \rangle_S = (4\pi \bar{R}^2)^{-1}\int_{S_0} d S$ denotes the surface average at a radial distance $r=\bar{R}$ corresponding to the outer edge of the boundary layer.

In general, the solvent particle of type $\alpha$ can interact with the catalytic and noncatalytic hemispheres through different potentials $W_{\alpha C}$ and $W_{\alpha N}$, respectively. A concentration gradient of product particles created by the reactions on the catalytic surface together with different interactions of the fuel and product species with the surface of the Janus motor give rise to a slip velocity at the outer edge of the boundary layer at $\bar{R}$. The value of the axisymmetric slip velocity can be computed using the diffusiophoretic mechanism,\cite{anderson:89,kapral:13} and is given by
\begin{equation}\label{eq:slip-vel-diff-W}
\mathbf{v}^{(s)}(\bar{R},\theta) = -\frac{k_BT}{\eta} \mbox{\boldmath $\nabla$}_{\theta} c_B(\bar{R},\theta) \big[\Lambda_N  + (\Lambda_C-\Lambda_N)\Theta(\theta)\big],
\end{equation}
where $\theta$ is the polar angle in a spherical polar coordinate system (see Fig.~\ref{fig:JP}(a)), $\mbox{\boldmath $\nabla$}_{\theta}$ is the gradient in the tangential direction, $c_B$ is the concentration of $B$ particles, $k_BT$ is the thermal energy at temperature $T$, $\eta$ is the viscosity of solvent, and $\Theta(\theta)$ is the characteristic function that is unity on the catalytic hemisphere ($ 0 < \theta < \pi/2$) and zero on the non-catalytic hemisphere ($ \pi/2 < \theta < \pi$). The  effects of interactions of the $A$ and $B$ particles with the Janus particle appear through the factors $\Lambda_C$ and $\Lambda_N$, where
\begin{equation}
\Lambda_{H} = \int_{0}^{\infty} dr\; r \left(e^{-\beta W_{BH}} - e^{-\beta W_{AH}}\right),
\end{equation}
with $H = C,N$. Here we assume that the species $\alpha$ interacts with the catalytic and noncatalytic hemispheres with the same potential, that is $\Lambda_C = \Lambda_N =\Lambda$, and Eq.~(\ref{eq:slip-vel-diff-W}) becomes
\begin{equation}\label{eq:slip-vel}
\mathbf{v}^{(s)}(\bar{R},\theta) = -\frac{k_BT}{\eta} \mbox{\boldmath $\nabla$}_{{\bf s}} c_B({\bf s}) \Lambda,
\end{equation}
and with hard potentials described in Eq.~(\ref{eq:hard_potential}) we have
\begin{equation}
\Lambda = \frac{1}{2} (R_{A}^2 - R_{B}^2).
\end{equation}

%To compute the net velocity of the Janus motor for our simple model, we assume that a species $\alpha$ interacts with the catalytic and noncatalytic hemispheres with the same potential, although the potential may differ for the interaction of the $A$ and $B$ species with the motor.~\cite{footnote:diff_interact} The axisymmetric slip velocity at $\bar{R}$ is given by,{\color{red}\cite{anderson:89,kapral:13}}
%\begin{equation}\label{eq:slip-vel}
%\mathbf{v}^{(s)}(\bar{R},\theta) = -\frac{k_BT}{\eta} (\mbox{\boldmath $\nabla$}_{{\bf s}} c_B({\bf s})) \Lambda,
%\end{equation}
%where $\theta$ is the polar angle in a spherical polar coordinate system, ${\bf s}$ denotes a point on a spherical surface at radial distance $\bar{R}$ from the motor, $c_B$ is the concentration of $B$ particles, and $k_BT$ is the thermal energy at temperature $T$. The effects of interactions of the $A$ and $B$ particles with the Janus particle appear through the factor,
%\begin{equation}
%\Lambda = \int_{0}^{\infty} dr\; r \left(e^{-\beta W_{BJ}} - e^{-\beta W_{AJ}}\right) = \frac{1}{2} (R_{A}^2 - R_{B}^2).
%\end{equation}

The concentration field that appears in Eq.~(\ref{eq:slip-vel}) may be determined from the solution of a reaction-diffusion equation. The form that this equation takes depends on how the system is maintained in a nonequilibrium state. Fuel $A$ may be supplied and product $B$ removed at distant boundaries or nonequilibrium reactions may occur in the bulk that catabolize product molecules and generate fuel, similar to the process that occurs in living cells. To model the latter case, we may assume a reaction of the form $B \stackrel{k_2}\rightarrow A$ in the bulk phase that serves to maintain the system in a nonequilibrium steady state. In the low P{\'{e}clet} number regime, the steady-state reaction diffusion equation with the bulk reaction can be written as $D\nabla^2 c_A(r,\theta) +k_2 c_B(r,\theta) = 0$. Since the total bulk concentration of the solvent particles satisfies $c_0 = c_A+c_B$, which we assume to hold locally, this equation may also be written as
\begin{equation}\label{eq:rd}
(\nabla^2 - \kappa^2)c_B(r,\theta)=0,
\end{equation}
where we have defined $\kappa^2=k_2/D$. The reaction-diffusion equation should be solved subject to the boundary conditions, $\lim_{r \to \infty}c_A(r,\theta)=c_0$, while the ``radiation" boundary condition~\cite{Collins_Kimball_49} on the Janus motor that accounts for the catalytic conversion of $A \to B$ on its surface is
\begin{equation}
k_D \bar{R} \partial_r c_A(r,\theta)|_{\bar{R}} = k_0 c_A(\bar{R},\theta) \Theta(\theta),
\label{eq:radialBC}
\end{equation}
where $k_0$ is the intrinsic reaction rate, $k_D = 4\pi \bar{R} D$ is the Smoluchowski rate constant for a diffusion controlled reaction.

 The solution of the reaction-diffusion equation~(\ref{eq:rd}) can be expressed as a series of Legendre polynomials,
\begin{equation}
c_B(r,\theta) = c_0 \sum_{\ell} a_{\ell} f_{\ell}(r) P_{\ell}(u),
\label{eq:cB}
\end{equation}
where $u=\cos\theta$. Substitution of Eq.~(\ref{eq:cB}) into Eq.~(\ref{eq:rd}) yields a Bessel equation for the radial function $ f_{\ell}(r)$ whose solution, subject to the boundary conditions given above, can written in terms of modified Bessel functions of second kind, $K_{\ell+\frac{1}{2}}(\kappa r) $, as
\begin{equation}
f_{\ell}(r) = \frac{K_{\ell+\frac{1}{2}}(\kappa r)}{\sqrt{\kappa r}} \frac{\sqrt{\kappa \bar{R}}}{K_{\ell+\frac{1}{2}}(\kappa \bar{R})}.
\end{equation}
The $a_\ell$ coefficients can be determined from the solution to a set of linear equations as, $a_{\ell} =  \sum_m ({\bf M}^{-1})_{\ell m} E_m$
where
\begin{eqnarray}
M_{\ell m} &=& \frac{2Q_{\ell}}{2\ell +1} \delta_{m\ell} + \frac{k_0}{k_D}\int_0^{1}du\:P_{m}(u)P_{\ell}(u), \nonumber \\
E_m &=& \frac{k_0}{k_D}\int_{0}^1 du\:P_{m}(u),
\end{eqnarray}
with $Q_{\ell} = \kappa \bar{R}\; K_{\ell+\frac{3}{2}}(\kappa \bar{R})/K_{\ell+\frac{1}{2}}(\kappa \bar{R}) - \ell$. The concentration profile in the absence of a bulk reaction ($k_2 = 0$) is recovered by taking the $\kappa \rightarrow 0$ limit of the equations above.  Note that $\lim_{\kappa \to 0} f_\ell(r) = (\bar{R}/r)^{\ell+1}$, corresponding to the solution of the reaction-diffusion system where fuel is supplied and product removed only at the distant boundaries of the system. Also note that $K_{\ell+\frac{1}{2}}(\kappa r)/\sqrt{\kappa r}\to e^{-\kappa r}/\kappa r$  for large $\kappa r$, which implies that the bulk reaction ``screens" the power law decay of the concentration field with the screening length $\kappa^{-1} = \sqrt{D/k_2}$ which determines the average distance that a product particle travels from the catalytic surface by diffusion before being converted back to a fuel particle. We shall use a bulk reaction to maintain the system out of equilibrium in the simulations presented below.

By taking the surface average of the slip velocity, these results may now be used to determine the Janus motor velocity, leading to
\begin{equation}
{V}_u = -\langle\hat{\pmb{u}}\cdot \mathbf{v}^{(s)}\rangle_S = \frac{k_BT}{\eta}\frac{2 c_0}{3 \bar{R}}a_1 \Lambda  = \frac{k_BT}{\eta}\frac{c_0}{3 \bar{R}}(R_{A}^2 - R_{B}^2)a_1.
\label{eq:Vu}
\end{equation}
Note that when the solvent particles individually interact with the same potential with the different hemispheres of the Janus motor as in this model, the motor velocity depends only on the $\ell = 1$ component of the concentration field due to the fact that the contributions from the surface average of other modes are zero. Also note that in the cases where $\Lambda_C \neq \Lambda_N$, one can see from the Eq.~(\ref{eq:slip-vel-diff-W}) the propulsion velocity will depend on the value of $\Lambda_C - \Lambda_N$.

\section{Simulation of Janus motor dynamics}\label{sec:micro-sim}
Consider a single Janus motor with radius $R$, mass $M$ and moment of inertia $I = \frac{2}{5} M R^2$ confined in a cubic box with linear size $L=50\:a_0$ and periodic boundaries. The simulation volume also contains $A$ and $B$ solvent particles with mass $m$ and total density $n_0$ at temperature $T$. In what follows, we use dimensionless units where mass is in units of $m$, lengths in units of $a_0$ and energies in units of $k_BT$. Time is then expressed in units of $t_0=({ma_0^2/k_BT})^{1/2}$.  In these units $R = 2.5$, $n_0=10$, $M = \frac{4}{3}\pi R^3 n_0 m \approx 655$ and $I  \approx 1636$.

Solvent particles interact with the Janus motor through modified bounce-back collisions as discussed earlier. In order to investigate the dependence of the propulsion velocity on the factor $\Lambda$, various combinations of collision radii, listed in Table~\ref{tab:forward}, were considered. Solvent particles interact among themselves through multiparticle collision dynamics (MPCD)~\cite{Malevanets_Kapral_99,Malevanets_Kapral_00,Kapral_08,Gompper_etal_09}, which combines effective multiparticle collisions at discrete time intervals $\tau = 0.1$ with streaming between two consecutive collisions, so that the mean free path is $\lambda = \tau ({k_BT/m})^{1/2}=0.1 $. Multiparticle collisions are carried out by first sorting the particles into cubic cells $\xi$ with linear size $a_0$. The postcollision velocity of particle $i$ in cell $\xi$ is given by
$\mathbf{v}_i' = \mathbf{V}_{\xi} + \hat{\mathcal{R}}(\mathbf{v}_i - \mathbf{V}_{\xi})$, where $\hat{\mathcal{R}}$ is a rotation matrix around a random unit axis by an angle $120^{\circ}$ and $\mathbf{V}_{\xi}$ is the center-of-mass velocity of all the solvent particles in the cell $\xi$.~\cite{footnote:grid_shift} In the streaming step, a solvent particle undergoes a bounce-back collision if it is moving toward and encounters the Janus motor as described in Sec.~\ref{sec:simple-model}, otherwise its position at next time step is $\mathbf{r}(t+\delta t) = \mathbf{r}(t) + \mathbf{v}(t) \delta t$, where $\delta t = 0.01$ is the time step size. With the parameters for the solvent given above, the solvent viscosity is $\eta = 7.93$ and the common self diffusion constant for the $A$ and $B$  solvent species is $D = 0.061$.

To maintain the system out of equilibrium, bulk reactions converting product $B$ particles back to fuel $A$ particles are carried out using reactive MPCD.~\cite{Rohlf_Fraser_Kapral_08} At each MPC collision step, the reaction, $B \stackrel{k_2}\rightarrow A$, takes place independently in each cell $\xi$ with probability $p^{\xi}(N_B^{\xi}) = 1 - e^{-a_2^{\xi}\tau}$, where $N_B^{\xi}$ is the total number of $B$ particles in cell $\xi$ and $a_2^{\xi} = k_2N_B^{\xi}$ with $k_2=0.01$ the bulk reaction rate. For our parameters the screening length is found to be $\kappa^{-1} \approx 2.5$, about the same as the radius of the Janus particle.

\section{Single Janus particle}\label{sec:one-janus}

A Janus motor can execute forward ($\Lambda > 0$), backward ($\Lambda<0$) directed motion or pure diffusive motion ($\Lambda=0$); also see Movies S1 and S2 for active motors.\dag First we investigate the purely diffusive dynamics of a Janus motor with collision radii $R_A=R_B=R$ so that $\Lambda=0$. The translational diffusion coefficient of the Janus motor, $D_J$, can be obtained from the long time behavior of the  mean squared displacement, $MSD(t) = 6D_J t$, and we find $D_J = 0.0028$, which is close to the Stokes-Einstein value, $D_J = k_BT/6\pi \eta R \simeq 0.0027$. The rotational diffusion coefficient, $D_r$, can be determined from the decay of the orientation correlation function, $\langle\hat{\pmb{u}}(t)\cdot \hat{\pmb{u}}(0)\rangle = \exp(-2D_r t)$. The simulation result is $D_r \simeq 0.00085$. The rotational diffusion coefficient is related to the rotational friction coefficient, $\zeta_r$, by $D_r=k_BT/\zeta_r$. The rotational friction coefficient can be expressed approximately in terms of microscopic and hydrodynamic contributions~\cite{HKW-77,Padding_Wysocki_Lowen_Louis_05,Whitmer_Luijten_10}, $\zeta_r = (\zeta_e^{-1} + \zeta_h^{-1})^{-1} \simeq 1417$, where $\zeta_e = \frac{8}{3}\sqrt{2\pi k_BT \mu} n_0 R^4 \big[2M/(5\mu + 2 M)\big]$ is the Enskog friction and $\zeta_h = 8\pi \eta R^3$ is the hydrodynamic friction for a spherical object. Using this expression for the rotational friction coefficient, we find $D_r \simeq 0.0007 $.

\begin{figure*}[htbp]
\centering
\resizebox{1.5\columnwidth}{!}{%
\includegraphics{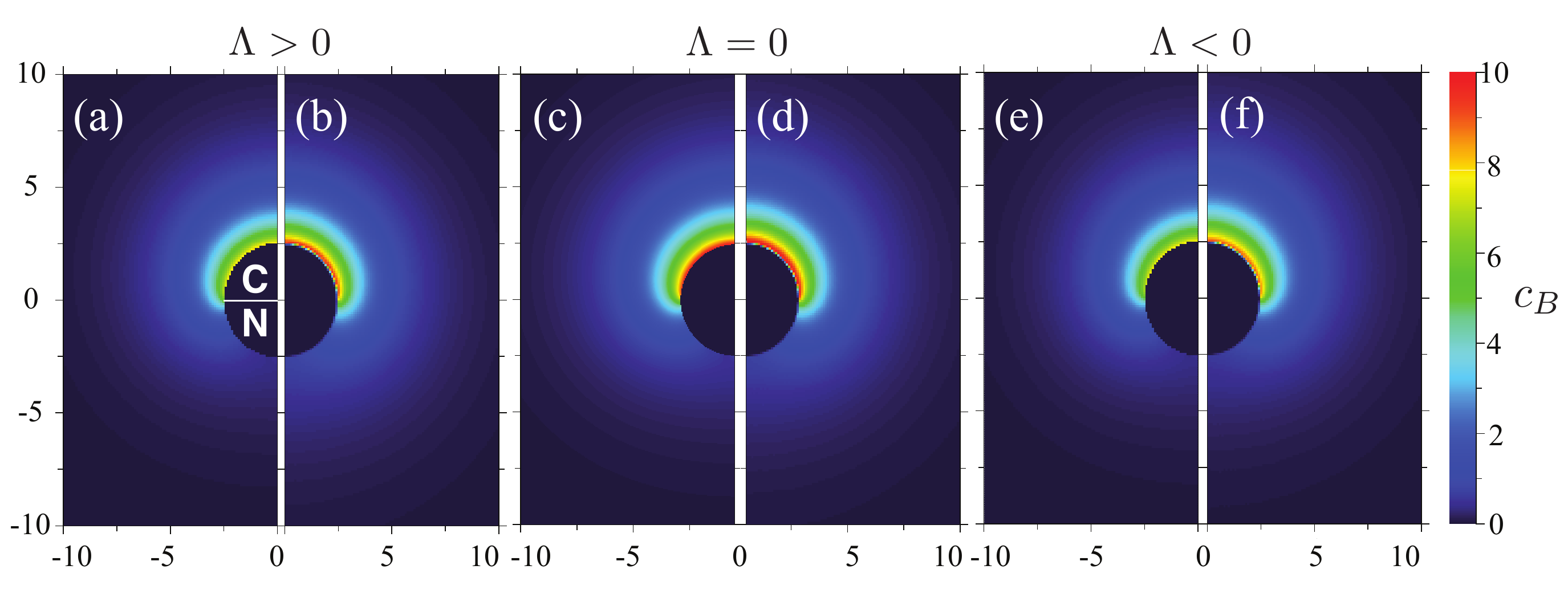}}
\caption{\label{fig:B_density_field} Product concentration field, $c_B(r,\theta)$, for moving ($|\Lambda| = 0.037$) and diffusive ($\Lambda = 0$) Janus motors. Panels (a), (c) and (e) are obtained from the analytical expression in Eq.~(\ref{eq:cB}), whereas (b), (d) and (f) are the simulation results. The catalytic and the noncatalytic hemispheres are labeled in panel (a).}
\end{figure*}

An active Janus motor will undergo directed motion along its symmetry axis as a result of chemically-powered propulsion, as well as translational and rotational Brownian motion. The simulation value of propulsion velocity of an active Janus motor may be determined from a time and ensemble average of its instantaneous velocity projected onto its instantaneous orientation, $V_u^S = \langle \mathbf{v}_J(t)\cdot \hat{\pmb{u}}(t)\rangle $, where $\langle ... \rangle$ denotes the average over time and realizations. Table~\ref{tab:forward} lists the average steady state propulsion velocity, $V_u^S$, for various values of $\Lambda$. As expected, the Janus particle switches from forward to backward motion when $\Lambda$ becomes negative, and its speed increases as $|\Lambda|$ increases. Note that for these propulsion velocities the P{\'{e}}clet number ($P_e = V_uR/D$) is $P_e < 0.4$.

\begin{table}
\small
\caption{\label{tab:forward}Properties of Janus particles with various $\Lambda$ factors: $k_0$, $k_f$ and $k_D = 4\pi D \bar{R}$ are the intrinsic, long-time and diffusion-controlled reaction rate coefficients, respectively; $\bar{R}$ is the radius of the outer edge of the boundary layer. $V_u^T$ and $V_u^S$ are the results of Janus particle velocity projected along particle axis $\hat{\pmb{u}}$ from theory and simulation, respectively. The numbers in parentheses are uncertain digits, $e.g.$, $1.23(4) = 1.23 \pm 0.04$.}
\begin{tabular}{llllll}
\hline \\
$R_A$&$2.5$&$2.5$&$2.5$&$2.485$&$2.47$\\
$R_B$&$2.47$&$2.485$&$2.5$&$2.5$&$2.5$\\
$\Lambda$&$0.075$&$0.037$&$0.0$&$-0.037$&$-0.075$\\
\\
$V_u^S$&$0.0090(3)$&$0.0043(3)$&$0.000(1)$&$-0.0044(3)$&$-0.0095(3)$\\
$V_u^T$&$0.012$&$0.0063$&$0.0$&$-0.0062$&$-0.013$\\
\\
$k_0$&$14(4)$&$15(4)$&$14(1)$&$15(4)$&$13(4)$\\
$k_f$&$1.75(2)$&$1.71(1)$&$1.7(2)$&$1.71(2)$&$1.69(1)$\\
$k_D$&$2.00$&$1.94$&$1.89$&$1.93$&$1.94$\\
$\bar{R}$&$2.60$&$2.52$&$2.46$&$2.52$&$2.52$\\
\\
$D_r$&$0.00087$&$0.00083$&$0.00085$&$0.00083$&$0.00084$\\
$\tau_r$&$576$&$604$&$586$&$600$&$597$\\
\\
$D_{e}^T$&$0.016$&$0.004$&$0.0028$&$0.004$&$0.018$\\
$D_{e}^S$&$0.02$&$0.006$&$0.0028$&$0.0036$&$0.02$\\
\end{tabular}
\end{table}

In order to compare these simulation results with the predictions of continuum theory, the intrinsic reaction rate coefficient $k_0$ and the location of the outer edge of the boundary layer $\bar{R}$ are needed to obtain the coefficient $a_1$ in Eq.~(\ref{eq:Vu}). The rate coefficient $k_0$, that governs the reaction $A \to B$ on the catalytic hemisphere of the Janus motor, can be computed in simulations by monitoring the time evolution of the total number of fuel $A$ particles in the system arising from the irreversible chemical reactions on the Janus particle.~\cite{Tucci_Kapral_04,Thakur_Kapral_11} In the low P{\'{e}}clet number limit the effects from the motion of the Janus particle can be neglected. The rate equation for $A$ particle concentration is $d c_A(t)/dt = -k_f(t) c_J c_A(t)$, where $c_J = 1/L^3 $ is the Janus motor number density. Here $k_f(t)$ is the time-dependent rate coefficient for the conversion of $A$ to $B$, which starts at $k_f(0^+) = k_0$ and decays to the asymptotic value $k_f = k_0 k_D/(k_0+k_D)$.~\cite{Tucci_Kapral_04} The outer edge of the boundary layer $\bar{R}$ can be defined as the distance within which the microscopic details of the dynamics of the interactions between solvent particles and the Janus particle become important so that a continuum description is not applicable. From the asymptotic value of $k_f = k_0 k_D/(k_0+k_D)$ we may determine $k_D$ and, use its value to determine $\bar{R}$.

The time-dependent rate coefficient, $k_f(t)$, was computed by measuring $-(d c_A(t)/dt)/(c_J c_A(t))$ in simulations that started with all fuel $A$ particles in the bulk of the solution and in the absence of the $B \to A$ bulk reaction ($k_2 = 0$). Table~\ref{tab:forward} shows the values of $k_0$ and $k_D$ extracted from the simulation data for various $\Lambda$ values, and the associated radius of outer edge of the boundary layer determined from $\bar{R} = k_D/4\pi D$. The intrinsic rate coefficient may be computed from a simple collision model. Since a reaction happens only when an $A$ particle is in contact with the collision surface at $R_A$, the rate $k_0$ is then given by the rate of collisions of the $A$ particle with the catalytic part of the Janus sphere, leading to $k_0 = {R_A^2} \sqrt{2\pi k_BT/m}$.\cite{footnote:k_0} It takes the value $k_0 \simeq 15.66$ for $R_A = 2.5$, which is slightly larger than the simulation values for various $\Lambda$. For the backward-moving Janus motors, as expected, $k_0$ decreases as $|\Lambda|$ increases because of the smaller collision radius $R_A$. Using Eq.~(\ref{eq:Vu}) with the parameters $k_0$, $k_D$ and $\bar{R}$, the computed theoretical propulsion velocities ($V_u^T$) for different values of $\Lambda$ are listed in Table~\ref{tab:forward}. We find the theoretical predictions are in good agreement with the simulation results.

\begin{figure}[htbp]
\centering
\resizebox{1.0\columnwidth}{!}{%
\includegraphics{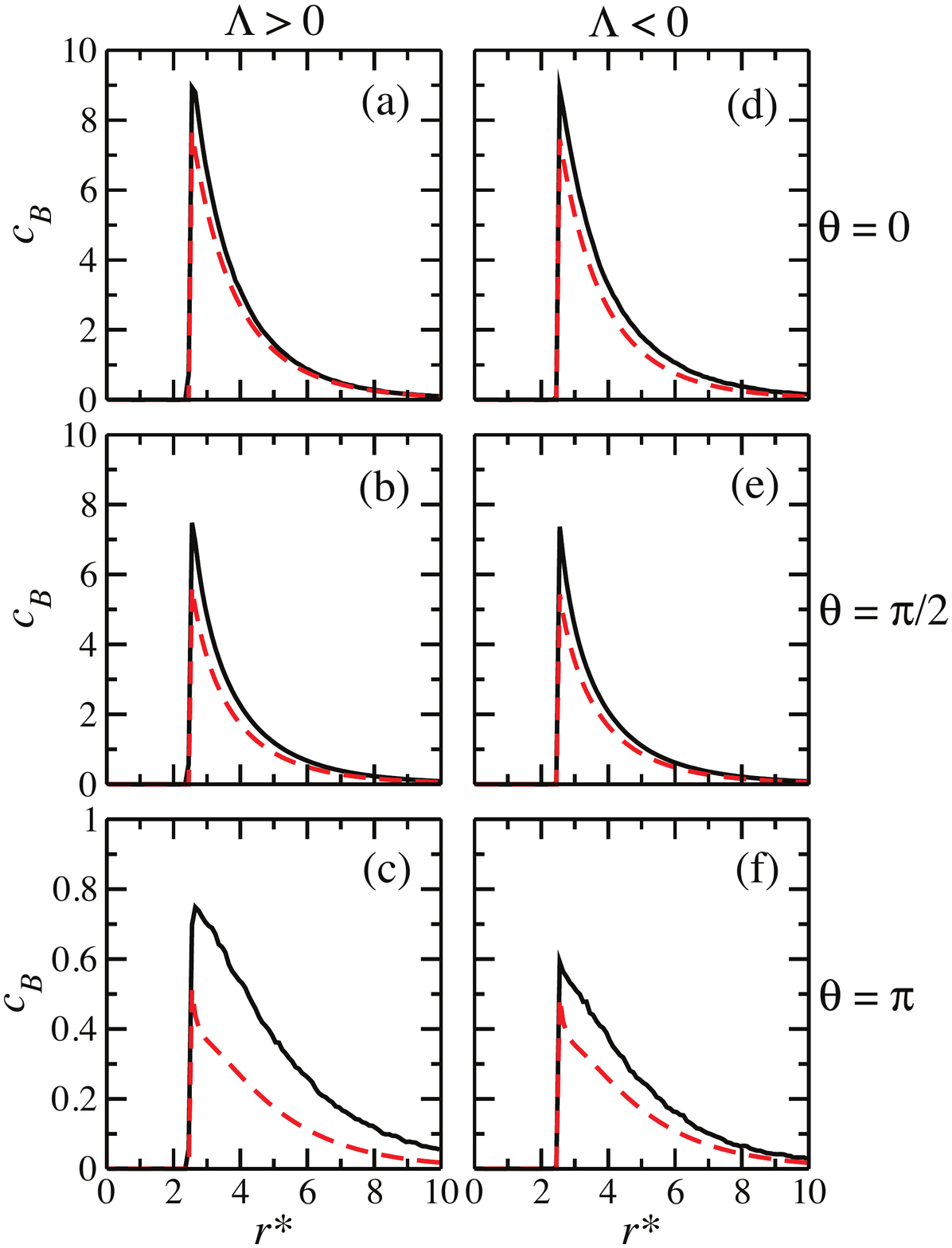}}
\caption{\label{fig:cB_compare}Comparison of product concentration profiles for the forward (a-c) and the backward (d-f) Janus particles with $|\Lambda| = 0.037$ obtained from simulation (black solid lines) and theory (red dashed lines) along various directions with $\theta = 0$, $\pi/2$, and $\pi$. Note that the ordinate scales on panels corresponding to different directions are not the same.}
\end{figure}

The steady-state concentration field of the product particle can be calculated analytically using Eq.~(\ref{eq:cB}), along with the coefficients $a_{\ell}$ derived in Sec.~\ref{sec:continuum-model} and using the reaction rates listed in Table~\ref{tab:forward}. Figure~\ref{fig:B_density_field} compares the analytical and simulation results outside the boundary layer for the forward-moving, backward-moving and the diffusive Janus motors. Quantitative comparisons of the product concentration field along various directions are shown in Fig.~\ref{fig:cB_compare}. From Fig.~\ref{fig:B_density_field}, one sees good agreement between the results at large distances from the Janus motor, with slightly higher product concentrations near the catalytic surface (also see Fig.~\ref{fig:cB_compare}). Such small deviations may be caused by perturbations induced by motor motion due to the fact that our simulations are in the small but finite P{\'{e}}clet number regime.

At short times Janus motors move ballistically with velocity ${V}_u\:\hat{\pmb{u}} $  as a result of their propulsion, but at long times their motion becomes diffusive with an enhanced diffusion constant given by $D_e = D_J + \frac{1}{3}V_u^2 \tau_r$,\cite{kapral:13} where $\tau_r = (2D_r)^{-1}$ is the characteristic time for the rotational diffusion. The rotational diffusion constants $D_r$ and reorientation times $\tau_r$ were measured for various values of $\Lambda$ and are listed in Table~\ref{tab:forward}. We find that the rotational dynamics is not affected by the directed motion of the Janus particle for both forward and backward propagation. The enhanced diffusion constant was also measured by a fit to the long-time values of the mean squared displacement. Good agreements between the simulation ($D_e^S$) and theoretical ($D_e^T$) estimates can be seen in Table~\ref{tab:forward}.

\section{Collective behavior of Janus motors}\label{sec:collective}
\begin{figure}[htbp]
\centering
\resizebox{0.8\columnwidth}{!}{%
\includegraphics{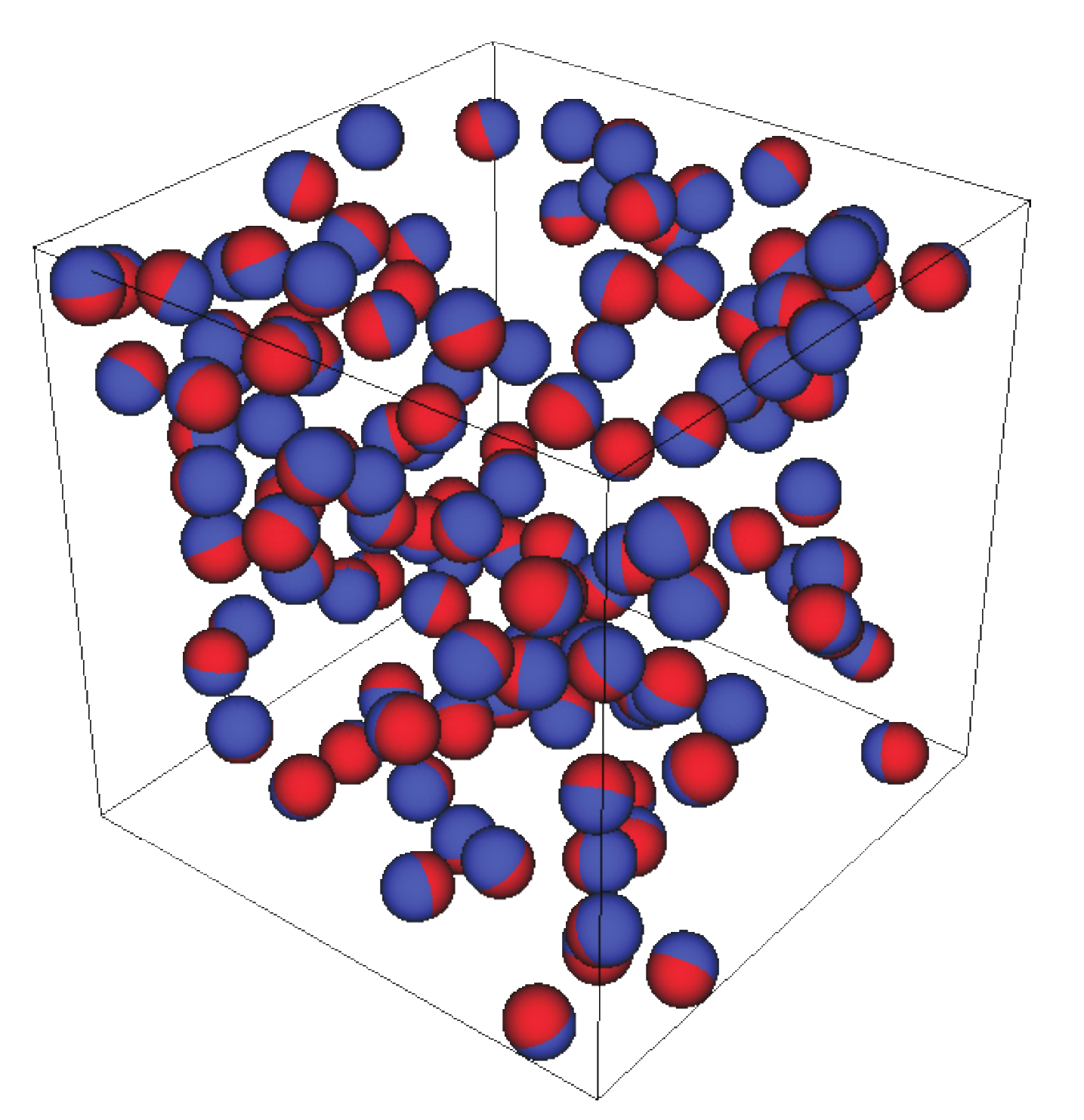}}
\caption{\label{fig:forward_collective}An instantaneous configuration taken from the simulation of a collection of forward-moving Janus motors with $\Lambda = 0.245$. A transient cluster can be seen in the upper-left corner of the simulation box. Also see Movies S3 and S4 for forward-moving and backward-moving Janus motors. The solvent particles are not displayed.\dag}
\end{figure}

The hard-sphere Janus motor model was also used to simulate the dynamics of a collection of Janus motors. In addition to the bounce-back collisions between solvent particles and the Janus motor, the interaction between any two Janus motors is described by a repulsive Lennard-Jones potential, $V_{JL}(r) = 4\epsilon [(\sigma/r)^{12} - (\sigma/r)^6 + 1/4]$, when their distance $r<2^{1/6}\:\sigma$. Here $\epsilon = 1$ is the interaction strength and $\sigma = 6$ is the effective radius, which is chosen to be larger than twice the hard-sphere radius $R$ so that each solvent particle can only interact with one Janus particle at a time. It is important to note that the attractive depletion force is negligible and does not dominate the collective dynamics. Simulations were carried out in a cubic box with linear size $L=50$ containing $N_J=125$ Janus particles, corresponding to a volume fraction of $\phi = \frac{4}{3}\pi R^3 N_J/L^3 \simeq 0.065$. A bulk reaction with $k_2 = 0.01$ was employed to maintain the system in a non-equilibrium steady state. Figure~\ref{fig:forward_collective} shows an instantaneous configuration of Janus motors taken from a realization of forward-moving Janus motors. The Janus motors are found to form transient clusters, which can be seen in the upper-left corner of the simulation box in Fig.~\ref{fig:forward_collective}. In contrast, no apparent clustering is observed for backward-moving Janus motors.

\begin{figure}[htbp]
\centering
\resizebox{1.0\columnwidth}{!}{%
\includegraphics{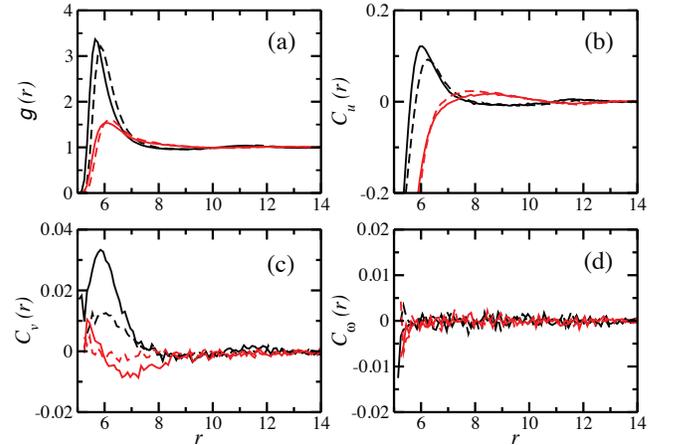}}
\caption{\label{fig:corre_fns}Spacial correlation functions: (a) Radial distribution function, $g(r)$, (b) Orientation correlation function, $C_u(r)$, (c) Velocity correlation function, $C_v(r)$, and (d) Angular velocity correlation function, $C_{\omega}(r)$, for $\Lambda = 0.48$ ($R_A = 2.5$ and $R_B = 2.3$, Black solid line), $\Lambda = 0.245$ ($R_A = 2.5$ and $R_B = 2.4$, Black dashed line), $\Lambda=-0.245$ ($R_A = 2.3$ and $R_B=2.5$, Red solid line), and $\Lambda = -0.48$ ($R_A = 2.4$ and $R_B=2.5$, Red dashed line).}
\end{figure}

\begin{figure}[htbp]
\centering
\resizebox{0.8\columnwidth}{!}{%
\includegraphics{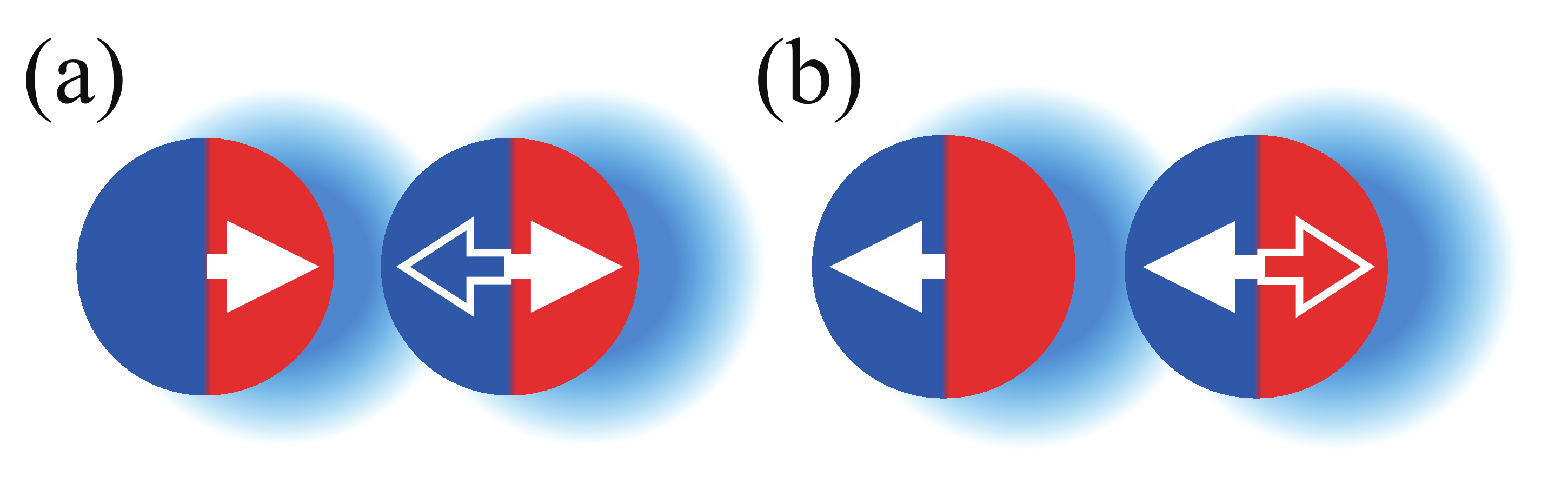}}
\caption{\label{fig:collect_align}Collective behavior of (a) two forward-moving and (b) two backward-moving Janus motors. The solid white arrows show the direction of motion of Janus motors driven by self-phoresis, whereas the hollow arrows indicate the direction of the force induced by the concentration gradient field (light blue clouds) generated by neighboring motors.}
\end{figure}

To quantitatively investigate the collective behavior of the forward and backward-moving Janus motors, first we consider the radial distribution function, $g(r)$, of the Janus motors,
\begin{equation}
g(r) =\bigg\langle  \frac{L^3}{4\pi r^2 N_J} \sum_{j<i=1}^{N_J} \delta\big(r_{ij} - r\big) \bigg\rangle,
\end{equation}
where $r_{ij} = |\mathbf{r}_{J, i} - \mathbf{r}_{J,j}|$ is the distance between motors $i$ and $j$. Fig.~\ref{fig:corre_fns}(a) shows $g(r)$ for various values of $\Lambda$. For negative $\Lambda = -0.245$ (red dashed line) and $-0.48$ (red solid line), there is a peak at the effective radial distance for the motor-motor repulsive interaction potential $r \approx \sigma$, and the peaks sharpen when $\Lambda = 0.245$ (black dashed line) and $0.48$ (black solid line) indicating the aggregation of forward moving Janus particles. Next, we study the steady-state properties of the collective motion by considering the correlation function,
\begin{equation}
C_{\beta}(r) =\bigg\langle  \frac{1}{n(r)} \sum_{j<i=1}^{N_J} (\hat{\beta}_i\cdot\hat{\beta}_j) \delta\big(r_{ij} - r\big) \bigg\rangle,
\end{equation}
where $\hat{\beta}_i$ is the unit vector of a physical quantity of particle $i$ and $n(r) = \sum_{j<i=1}^{N_J} \delta\big(r_{ij} - r\big)$ is the number of particle pairs with separation $r$. Figure~\ref{fig:corre_fns}(b) shows the orientational correlation functions, $C_u(r)$, for various values of $\Lambda$. While there is no significant correlation among the Janus particles with $\Lambda < 0$, when $\Lambda >0$ a positive peak found at $r=\sigma$, suggesting orientational alignment for forward-moving Janus particles. While several factors, such as induced flow fields and crowding effects, may affect the orientation alignment indicated in this figure, the interactions mediated by product concentration fields play very important roles in determining the dynamics of a collection of diffusiophoretic motors.\cite{Huang_Kapral_16} Such effects are the strongest when two motors are in an aligned configuration as shown in Fig.~\ref{fig:collect_align} (a) and (b) for the forward-moving and the backward-moving Janus motors, respectively. A single forward Janus motor propels itself toward the region with higher product concentration due to self-diffusiophoretic mechanism. In Fig.~\ref{fig:collect_align} (a), when two motors with orientational alignment are close to each other, the right motor feels the product concentration field generated by the left motor. Consequently, the motor on the right side tends to move toward the catalytic face of the left motor which stabilizes this oriented configuration. In contrast, a backward-moving motor prefers to move away from the regions with high product concentration and, therefore, the configuration shown in Fig.~\ref{fig:collect_align} (b) is unstable giving rise to lower orientational order. In addition, from the velocity correlation functions, $C_v(r)$, one can see that the forward-moving particles not only align but also propagate in the same direction as indicated by the positive peak at $r=\sigma$. For the backward-moving particles, a broad negative peak was found at $r = 7$ showing that two particles move away from each other. Finally, we compute the angular velocity correlation function, $C_{\omega}(r)$. The Janus particles interact with each other through central potentials and thus no angular momentum exchange happens during elastic collisions. Therefore, as expected no significant angular velocity correlation among the Janus particles was found.

\section{Discussion}\label{sec:conc}
Systems with active elements occur throughout nature and are currently being investigated in many laboratories. The focus of these investigations varies, ranging from studies of cargo transport involving single motors to the more complex dynamics of many interacting motors. As the systems under investigation become more complicated, for example, involving many interacting motors to study nonequilibrium phase transitions or phase segregation, or crowded systems with mobile obstacles of arbitrary shape, molecular simulation provides a promising way to discover the essential features that underlie the physical phenomena and to predict what new phenomena might be seen.

Continuum models for phoretic propulsion are certainly applicable to large motors and, in fact, often provide good results for small submicron scale motors. They will breakdown on the smallest scales and for the smallest motors. The results in this paper provide some insight into how the parameters that enter into continuum models may be determined in order to make comparisons with simulations of small motors in fluctuating molecular environments.

For many-motor systems microscopic dynamics that satisfies the basic conservation laws of mass, momentum and energy will correctly account for all aspects of coupling that arise from hydrodynamic flow fields induced by motor motion, concentration gradients that have their origin in the catalytic activity of all motors, as well as direct motor-motor interactions. Effects, such as those arising from variations of an individual motor's speed due to perturbations of chemical gradients by other motors in the system and the chemotactic-like interactions due to these gradients, are incorporated in the simulations. The simplicity of the Janus model described in this paper will facilitate large-scale simulations designed to probe collective behavior, beyond the illustrative examples presented in the text. More generally, microscopic models will provide a way to analyze the delicate interplay of effects that contribute to the new phenomena that are being explored in chemically-active motor systems.

\section*{Acknowledgements}
\label{sec:acknowledge}
This work was supported by a grant from the Natural Sciences and Engineering Research Council of Canada. Computations were performed on the GPC supercomputer at the SciNet HPC Consortium. SciNet is funded by: the Canada Foundation for Innovation under the auspices of Compute Canada; the Government of Ontario; Ontario Research Fund - Research Excellence; and the University of Toronto.~\cite{scinet}

%%%END OF MAIN TEXT%%%

%The \balance command can be used to balance the columns on the final page if desired. It should be placed anywhere within the first column of the last page.

%\balance

%If notes are included in your references you can change the title from 'References' to 'Notes and references' using the following command:
\renewcommand\refname{Notes and references}

%%%REFERENCES%%%
%\bibliography{reference} %You need to replace "rsc" on this line with the name of your .bib file

\begin{thebibliography}{48}
\bibitem [{foo({\natexlab{b}})}]{footnote:k_0}%
  \BibitemOpen
  \href@noop {} {} ({\natexlab{b}}),\ \bibinfo {note} {the intrinsic rate
  constant $k_0$ may be estimated from simple collision theory as the product
  of the mean velocity of solvent particles $\bar{v} = \sqrt{8 k_BT / \pi m}$
  times the cross section $\sigma_0 = \pi R_A^2$, $k_0=\frac{1}{2}\bar{v}
  \sigma_0$, where the prefactor $1/2$ accounts for the fact that only
  collisions with the catalytic hemisphere lead to reaction.}\BibitemShut
  {Stop}%
\end{thebibliography}
%\bibliographystyle{rsc} %the RSC's .bst file
%merlin.mbs apsrev4-1.bst 2010-07-25 4.21a (PWD, AO, DPC) hacked
%Control: key (0)
%Control: author (8) initials jnrlst
%Control: editor formatted (1) identically to author
%Control: production of article title (-1) disabled
%Control: page (0) single
%Control: year (1) truncated
%Control: production of eprint (0) enabled
\providecommand{\noopsort}[1]{}\providecommand{\singleletter}[1]{#1}

\end{document}